\documentclass[runningheads]{llncs}

\usepackage{graphicx}
\usepackage{xspace}
\usepackage{subcaption}
\usepackage{color}
\usepackage{xcolor}
\usepackage{comment}
\usepackage[hyphens]{url}
\usepackage{hyperref}
\usepackage[absolute,overlay]{textpos}

\newcommand{\ie}{\mbox{i.e.,}\xspace}
\newcommand{\eg}{\mbox{e.g.,}\xspace}

\newcommand{\WP}{\mbox{WP}\xspace}
\newcommand{\pornhubreview}{\url{www.pornhub.com/insights/2017-year-in-review}}

\newcommand{\lu}[1]{{\color{violet}{[lv: #1]}}}
\newcommand{\lv}[1]{{\color{violet}{[lv: #1]}}}

\begin{document}

\title{Characterizing web pornography consumption from passive measurements}

\author{
    Andrea Morichetta, Martino Trevisan, Luca Vassio \\
    Politecnico di Torino \\
    \texttt{\small first.last@polito.it}
    \thanks{The research leading to these results has been funded by the Vienna Science and Technology Fund (WWTF) through project ICT15-129 (BigDAMA) and the Smart-Data@PoliTO center for Big Data technologies.}
}

\institute{}




\maketitle 

\TPshowboxestrue
\TPMargin{0.3cm}
\begin{textblock*}{18cm}(1.8cm,0.7cm)
\bf
\definecolor{myRed}{rgb}{0.55,0,0}
\color{myRed}
\noindent
Please cite this paper as:
Morichetta A., Trevisan M., Vassio L. (2019) Characterizing Web Pornography Consumption from Passive Measurements. In: Choffnes D., Barcellos M. (eds) Passive and Active Measurement. PAM 2019. Lecture Notes in Computer Science, vol 11419. Springer, Cham. \url{https://doi.org/10.1007/978-3-030-15986-3\_20}
\end{textblock*}

\begin{abstract}

Web pornography represents a large fraction of the Internet traffic, with thousands of websites and millions of users. Studying web pornography consumption allows understanding human behaviors and it is crucial for medical and psychological research. 
However, given the lack of public data, these works typically build on surveys, limited by different factors, \eg unreliable answers that volunteers may (involuntarily) provide.

In this work, we collect anonymized accesses to pornography websites using HTTP-level passive traces. Our dataset includes about $15\,000$ broadband subscribers over a period of 3 years. We use it to provide quantitative information about the interactions of  users with pornographic websites, focusing on time and frequency of use, habits, and trends. We distribute our anonymized dataset to the community to ease reproducibility and allow further studies.

\end{abstract}

\section{Introduction}
\label{sec:intro}

Pornography and technology have enjoyed a close relationship in the last decades, with technology hugely increasing the capabilities of the porn industry. From the limited market reachable through public theatres, the introduction of the videocassette recorder in the 1970s 
abruptly changed the way of accessing pornography, 
allowing to access pornography in the privacy and comfort of each individual home. Then, the birth of cable networks and specialty channels in the 1990s, allowed a further step towards accessibility and privacy, giving the possibility to retrieve content directly from home. 
Finally, the Internet revolutionized again the market, guaranteeing direct desktop delivery to every individual with a connection, interactivity through forums and webcams, free content and, at the same time, anonymity. In 2017, the most used pornographic platform in the world (Pornhub, according to Alexa ranking)\footnote{\label{notaAlexa}As of October $17^{th}$, 2018 \url{www.alexa.com/topsites}}, claims 80 million daily accesses to its website.\footnote{\label{note1}\pornhubreview} The importance of Internet pornography as a prevalent component of popular culture and the need of its study has been recognized for a long time \cite{DILEVKO200436}.

In this work we do not attempt to classify Internet pornography. Rather, we refer to the term web pornography (\WP) to any online material that, directly or indirectly, \emph{seeks to bring about sexual stimulation}~\cite{librovecchio}. Therefore, the term \emph{pornographic website} is here used to describe services that provide actual pornographic videos, sell sex related merchandise, help in arranging sexual encounters, etc. 
We refer only to adult pornography websites and we do not advocate the inclusion of child pornography websites in our research, thus this paper has no application whatsoever to child pornography. The word pornography, in the context of this article, refers exclusively to legal content in the territories of EU and USA.

Through the years, \WP has been the subject of many studies, that aimed at describing how people make use of it, or pinpointing eventual pathological situation correlated to an excessive use. However, such works typically come from the medical and psychology communities, and are based on surveys that cover a very small number of volunteers. Moreover, previous studies~\cite{SurveyLiars1,SurveyLiars2} report that people tend to lie, either consciously or unconsciously, when answering to private-life concerning surveys, especially about sexuality; there are people who declare more accesses than real (\eg to show to be uninhibited) and others who understate the actual consumption, fearing social blame.  
Both these behaviours, called social desirability biases, and egosyntonic/egodystonic feelings (\ie being or not in accordance with their self-image) make surveys less reliable with respect to other sources of information.

In contrast to previous works, in this study we investigate \WP by means of passive network measurements, collected from about $15\,000$ broadband subscribers over a period of 3 years.  MindGeek, a company operating many popular pornographic websites, switched to encryption only in April 2017, being the first big player of \WP industry to adopt HTTPS
.\footnote{\url{www.washingtonpost.com/news/the-switch/wp/2017/03/30/porn-websites-beef-up-privacy-protections-days-after-congress-voted-to-let-isps-share-your-web-history}} As such, the vast majority of \WP portals used plain-text HTTP up to March 2017, allowing us to leverage HTTP-level measurements, and obtain detailed results of \WP consumption. Using recent advances in data science, we extract only user actions to \WP portals from a deluge of HTTP data.


The main contributions of this paper are:
\begin{itemize}
\item We provide a thorough characterization of \WP consumption leveraging measurements from $15\,000$ broadband subscribers over a period of 3 years.
\item We show how users moved to mobile devices through the years, even if the time spent on\WP remains constant.
\item We show that typical \WP sessions last less than 15 minutes, with users rarely accessing more than one website. Less than 10\% of users consume \WP more than 15 days in a month, and repeated use within a single day is very sporadic.
\item We release our dataset to the community in anonymized form for further investigation~\cite{our_dataset}. To the best of our knowledge, this is the only public datasets that includes \WP accesses from regular Internet users.
\end{itemize}

The employed metrics are taken from the surveys reported in medical literature and from \WP portal reports that we use throughout the paper. We restrict our analysis only to those that we were able to verify given our data.
Our results enhance the visibility and understanding of those topics, and give a less mediated overview of users behaviors, mostly confirming what emerges from medical surveys. 

The remainder of the paper is organized as follows: Section~\ref{sec:related} summarizes related work. Section~\ref{sec:measurements} describes data collection, processing and privacy issues, while Section~\ref{sec:results} presents the results. Finally, Section~\ref{sec:conclusion} concludes the paper.

\section{Related Work}
\label{sec:related}

Most previous works that investigate the interaction between users and \WP leverage the information contained in surveys proposed to groups of volunteers. 
Vaillancourt-Morel~\emph{et~al.}~\cite{vaillancourt_profiles_2017} examine the potential presence of different profiles of pornography users and their relation with sexual satisfaction and sexual dysfunction. The investigation is conducted over a poll that involved 830 adults, and they group users' behavior in three clusters. 
Daspe~\emph{et~al.}~\cite{daspe_when_porn_2018} investigate the relationship between frequency of \WP consumption and the personal perception of this behavior, pointing out that often there are strong discrepancies. Another analysis of the phenomenon is provided by Grubbs~\emph{et al.}~\cite{grubbs_moral_2014}, where the analysis is conducted over two participants sets, divided in students and adults, showing that moral scruples can infect the self-impression over their consumption.
Wetterneck~\emph{et~al.}~\cite{short_review_2012} propose a critical analysis of \WP, showing the various limitations of the state of the art of studies that assessed online pornography usage, concerning its definition, consumption, and the variability of its measurements.

Fewer works used network measurements to study \WP. 
Tyson \emph{et~al.}~\cite{uhlig_porn_2016} extract trends and characteristics in a major adult video portal (YouPorn) by analyzing almost 200k videos, together with meta-data such as page content, ratings and tags.
In a similar direction, Mazieres~\emph{et~al.}~\cite{mazieres_deeptag_2014} produce and analyze a semantic network of \WP categories, extracted from the portal xHamster, in order to find which are the most dominants and if they are actually meaningful.
Ortiz~\emph{et~al.}~\cite{ortiz_2005_characterizing} study a Chilean websites containing human images and classify them in normal, porno and nude, with the objective of automatically discovering \WP websites.
Finally, Coletto~\emph{et~al.}~\cite{coletto_2016_pornography} study users' activity in social networks related to \WP, in order to extract information about the seclusion of those communities with respect to the rest of the population and their characteristics in terms of age and habits and gender.
To the best of our knowledge, we are the first to use passive measurements to study the behavior of users accessing web pornography.

\section{Measurements}
\label{sec:measurements}

\subsection{Data Collection}
In this work, we rely on network measurements coming from passive monitoring of a population of broadband subscribers over a period of 3 years (from March 2014 to March 2017). We have instrumented a Point-of-Presence (PoP) of a European ISP, where $\approx 10\,000$ ADSL and $\approx 5\,000$ FTTH customers are aggregated. ADSL downlink capacity is 4--20 Mbit/s, with uplink limited to 1 Mb/s. FTTH users enjoy 100 Mb/s downlink, and 10 Mbit/s uplink. Each subscription refers to an installation, where users’ devices (PCs, smartphones, etc.) connect via WiFi or Ethernet through a home gateway. Important to our analysis, the ISP provides each customer a fixed IP address, allowing us to track her over time. Nevertheless, a small fraction of customers abandoned the ISP during the observation period, and few new ones joined. All ADSL customers are residential customers (i.e., households), while a small number of business customers exist among the FTTH customers.

To gather measurements we use Tstat~\cite{tstat_dpdk}, a passive meter that collects rich per-flow summaries, with hundreds of statistics regarding TCP/UDP connections issued by clients. Beside, Tstat includes a DPI module that creates log files containing details about observed HTTP transactions. For each transaction, it records the URL, a client identifier as well as other HTTP headers of requests and responses.
Our measurements are based on the inspection of HTTP headers, and, as such, neglect all encrypted traffic. However, no big \WP portal used encryption at the time our dataset was collected.
Generated log files are copied to our back-end servers with a daily frequency. Data is stored on a medium-sized Hadoop cluster to allow scalable processing. All processing is done using Apache Spark and Python. The stored data covers 3 years of measurements, totaling 20.5 TB of compressed and anonymized flow logs (around 138 billion records).

\subsection{Definition of user and its limitations}
\label{sec:definition_user}

Our PoP is located at the Broadband Remote Access Server (BRAS) level. Each subscription is identified by a unique and fixed IP address. However, subscriptions typically refer to households where potentially more than one person surf the Internet sharing the same public IP address. As such, relying on the client IP to identify a user would not be precise enough to study habits and behavior. Thus, in our work we define a \emph{user} as the concatenation of the client IP address and the user-agent as extracted from the corresponding HTTP header. Note that with this definition a single person may appear multiple times with different identifiers if she uses multiple devices or her device incurs software updates that modify the user agent string. Analyses are thus performed on a per-browser fashion -- i.e., each user-agent string observed in a household. Privacy requirements limit any finer granularity.


The evaluated dataset includes only a regional sample of households in a single country. Users in other regions may have diverse browsing habits. Equally, mobile devices have been monitored only while connected to home WiFi networks. As such,
our quantification of browsing on mobile terminals is actually a lower-bound, since visits while
connected on mobile networks are not captured.

\subsection{Data filtering and session definition}

Starting from a HTTP-level dataset, we need to filter only entries referring to \WP websites. Studying innovative methodologies to automatically isolate traffic towards particular services is out of the scope of this work. We employ a blacklist based approach to perform classification. We build on public available lists, achieving robustness by combining three different sources.\footnote{\url{www.shallalist.de/categories.html},   \url{www.similarweb.com}, and \\
\url{dsi.ut-capitole.fr/blacklists/index_en.php}}
These three lists provide a set of domain names that offer different \WP content (ranging from video streaming to thematic forums). To avoid false positives, we consider only those domain names contained in at least two over three lists. We come up with $310\,252$ unique entries, arranged over $460$ top-level domains.

After filtering entries referring to \WP websites, we perform a further step to identify \emph{sessions} of continuous activity. To this end, we group data by user, and process HTTP transactions by start time.
We then identify session as follows: when a user accesses a pornographic website we open a new session and account to this all subsequent entries to \WP websites. We terminate a session if we do not observe any entry to \WP for a period of $30$ minutes. 
While defining a browsing session is complicated~\cite{fomitchev2010google}, we simply consider a time larger than 30~minutes as an indication of the session end as it is often seen in previous works (\eg \cite{Catledge1995}), and in applications like Google Analytics.\footnote{\url{support.google.com/analytics/answer/2731565?}}

\subsection{User actions extraction}

Subsequently, we further filter the dataset to isolate only those HTTP requests containing an explicit user action by the user. This step aims at isolating users' behavior
discarding all HTTP traffic related to inner objects of webpages such as images, style-sheets, and scripts
To this end, we implement the methodology described in our previous work~\cite{vassio_2016_detecting} that builds a machine-learning model to pinpoint intentionally visited URLs (\ie webpages) from raw HTTP traces.
The followed strategy has as core module a supervised classifier, which is able to correctly recognize user actions in HTTP traces. It results to reach an accuracy of over 98\%, and it can be successfully applied to different scenarios, including smartphone apps~\cite{Vassio:2018}. 

In total, after the extraction, we have 58 million user actions/visited webpages towards 59\,989 different adult domains. We observe an average of 13\,261 different \WP users per month. For each user, we determine information about used OS, browser,  
and if the device was a PC, a smartphone, or a tablet. These information are extracted from the user-agent of the original HTTP request at the time of the capture, using the Universal Device Detection library.\footnote{\url{github.com/piwik/device-detector}}
We made these data available to the community in anonymized form to guarantee reproducibility of our results and further investigations~\cite{our_dataset}. In the remainder of the paper, we only take into account user actions, to which we simply refer with the term \emph{visited webpages}.


\subsection{Privacy and ethical concerns}

Passive measurements potentially expose information which may threaten users’ privacy. As such, our data collection program has been approved by the partner ISP and by the ethical board of our University. Moreover, this specific data analysis project was also subject to a privacy impact assessment that was done with the data protection officer of our institution.

We undertake several countermeasures to avoid recording any personally identifiable information. Before any storage, all client identifiers are anonymized using Crypto-PAn algorithm~\cite{fan2004crypto}, and URLs are truncated to avoid recording URL-encoded parameters. Encryption keys are varied on a monthly basis, to avoid persistent users tracking.
Sensitive information such as cookies and \texttt{Post} data are not monitored at all. Logs are stored in a secured data center in an encrypted format.
We emphasize again that in our research we only refer to adult pornography websites, obtained through open datasets, referring exclusively to legal content in the territories of EU and USA.

\section{Results}
\label{sec:results}

In this section, we report the most significant results emerging from our dataset. We first focus on the time dimension, showing the evolution of \WP consumption from 2014 to 2017 in terms of quantity and device type. We then focus on users, characterizing duration and frequency of their \WP use. Finally, we provide some figures about the popularity of services.

\subsection{Usage trends}

\begin{figure}[!t]
    \begin{center}
        \begin{subfigure}{0.725\textwidth}
            \includegraphics[width=\columnwidth]{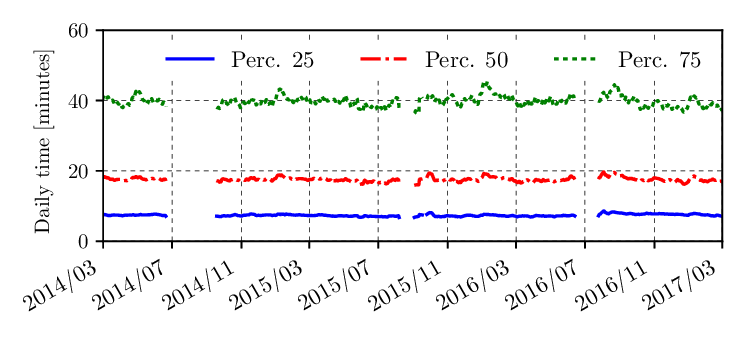}
             \caption{Per-user daily time spent on \WP. $25^{th}$, $50^{th}$, and  $75^{th}$ percentiles are shown.}
            \label{fig:duration_time}
        \end{subfigure}\\
        \begin{subfigure}{0.725\textwidth}
            \includegraphics[width=\columnwidth]{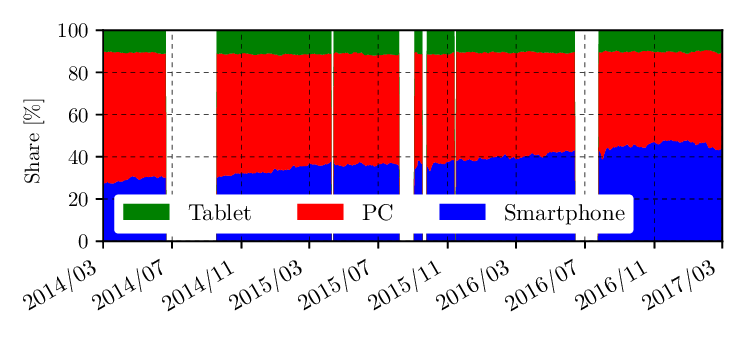}
            \caption{Device category share for accessing \WP.}
           \label{fig:ua_share}
        \end{subfigure}
        \caption{Usage trends from March 2014 to March 2017.}
        \label{fig:trends}
    \end{center}
\end{figure}

Our first analysis aims at describing \WP consumption trends from 2014 to 2017. 
In Fig.~\ref{fig:duration_time}, we focus on the time spent on \WP by monitored users. The blue (solid), red (dash-dot) and green (dashed) curves report, respectively, the $25^{th}$, $50^{th}$ and $75^{th}$ percentiles of the total per-user daily time spent on \WP, i.e., the sum of the duration of all the \WP sessions. Curves are calculated only for active users, i.e., users visiting at least one \WP website during one day. Curves are not continuous, for the lack of data due to outages in out PoP.  The outcome shows a rather stable trend over the observation period, with half of the users spending less than 18 minutes per day on \WP; however almost 25\% of users reaches 40 minutes of daily activity. These are day-wise statistics, and do not provide figures about the repeated use of \WP across multiple days by the same user, as we will see later.
Measuring the overall share of users accessing \WP portals is not easy using our data, as a single identifier -- the client IP address -- identifies a broadband subscription, potentially shared by multiple users. However, we notice that every day 12\% of subscribers access \WP websites, and this value is constant across years. A further analysis on \WP pervasiveness is given in Sec.~\ref{sec:sessions}.  

Those results can be used as a comparison with surveys statistics, fortifying or confuting what the participants declare. Vaillancourt-Morel \emph{et al.}~\cite{vaillancourt_profiles_2017} study the characteristics of \WP consumers. The majority of the chosen sample uses \WP for recreation only, on average 24 minutes per week, a value consistent but slightly higher compared to our data.

Then, we investigate the evolution in device categories use (PCs, tablets and smartphones). We compute, for each device category, its share in terms of number of sessions.
Fig.~\ref{fig:ua_share} shows the results.
We notice that
smartphones (blue surface), have largely increased their share from 27\% to 42\% at the expense of PCs. 
Tablets pervasiveness, reported in green, is instead rather constant. Not shown, the evolution of daily time spent with different devices did not changed too much throughout the years (see Sec.~\ref{sec:sessions} for more details).


\subsection{Usage frequency, duration and habits}
\label{sec:sessions}

\begin{figure}[!t]
    \begin{center}
    
        \begin{subfigure}{0.49\textwidth}
            \includegraphics[width=\columnwidth]{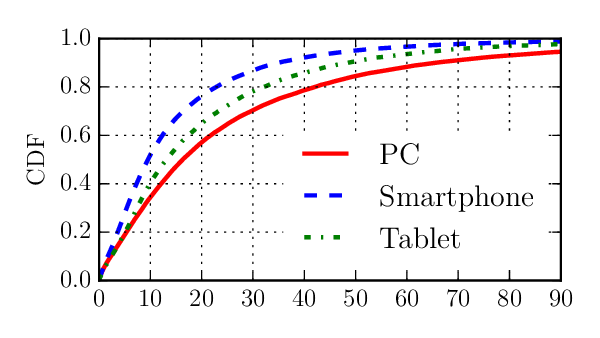}

             \caption{Session duration [minutes].}
            \label{fig:duration}
        \end{subfigure}
        \begin{subfigure}{0.49\textwidth}
            \includegraphics[width=\columnwidth]{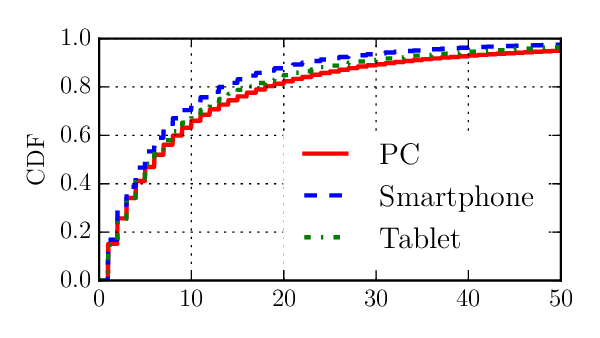}
            \caption{Accessed webpages per session [-].}
            \label{fig:clicks}
        \end{subfigure}
        \begin{subfigure}{0.49\textwidth}
            \includegraphics[width=\columnwidth]{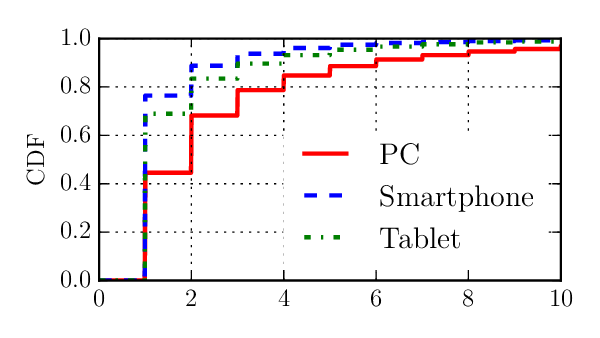}
            \caption{Accessed websites per session [-].}
           \label{fig:n_websites}
        \end{subfigure}
        \begin{subfigure}{0.49\textwidth}
            \includegraphics[width=\columnwidth]{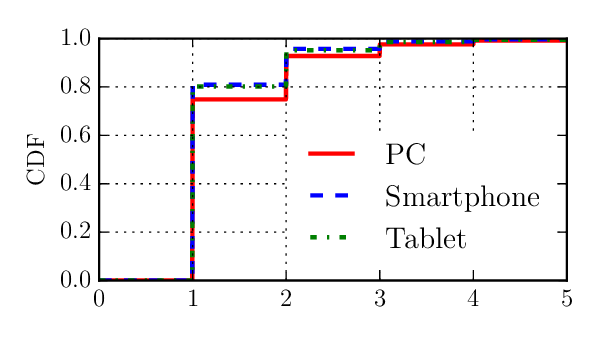}

            \caption{Sessions per day for an active user [-].}
           \label{fig:sessions_per_day}
        \end{subfigure}
        \caption{CDF of \WP session characteristics, divided by device type.}
        \label{fig:usage}
    \end{center}
\end{figure}

For detailed analyses, we restrict to the last month of our dataset that does not include nor public holidays nor measurement outages, \ie October 2016.
We first characterize \WP sessions in terms of duration.
Fig.~\ref{fig:duration} shows the empirical cumulative distribution function (CDF) of session duration, expressed in minutes. The duration is larger for PCs than for tablet and smartphones. While most of the sessions are rather short, i.e., less than 15 minutes for PCs and 10 for smartphones, we observe sporadic longer sessions up to more than one hour. 
We now draw the attention
on the number of webpages accessed during \WP sessions, whose CDF is reported in Fig.~\ref{fig:clicks}. 
Here the difference among devices is limited, with users accessing in median 5 or 6 webpages in a session, with 28\% of them limited to one or two. 
However, in some cases tens of webpages are accessed. 
Similarly, in Fig.~\ref{fig:n_websites} we report the distribution of the number of \emph{unique} websites accessed during a \WP session. 
Results show that smartphone users tend to focus on a single \WP website at a time (78\% of sessions), while PC users are more prone to visit more websites. For all the devices, very few sessions include visits to 4 or more different websites.
Finally, Fig.~\ref{fig:sessions_per_day} reports the number of daily sessions for an active user. 
The figure shows that users hardly make repeated use of \WP within a day, without differences among devices. 

\begin{figure}[!t]
    \begin{center}
        \includegraphics[width=0.75\textwidth]{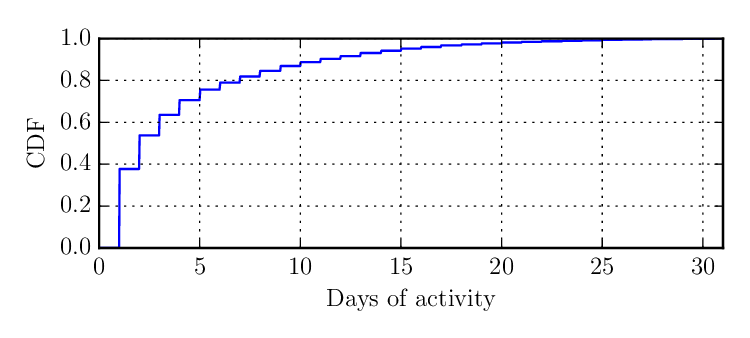}
        \caption{Number of distinct days in which users consumed \WP in a month.}
        \label{fig:cdf_days_porn}
    \end{center}
\end{figure}

We next focus on the frequency of \WP consumption by users over the month.
In Fig.~\ref{fig:cdf_days_porn} we report the CDF of number of days of activity for \WP users in the dataset. The figure indicates that the monthly frequency is generally low, with 76\% of the users visiting \WP 5 or less days in a month. Still, there are some users with a reiterate usage, with 8\% of them consuming \WP more than 15 days. These results confirm what is found by Daspe~\emph{et al.}~\cite{daspe_when_porn_2018}, who show that the 73\% of the participants to a survey access pornography no more than once or twice per week, and only 11\% more than 5 times per week. 
Given the nature of our dataset, we cannot estimate the number of users \emph{not} consuming \WP. Still, an analysis of per-\emph{subscription} traffic to \WP is provided later in this section.

\begin{figure}[!t]
    \begin{center}
    \includegraphics[width=0.77\textwidth]{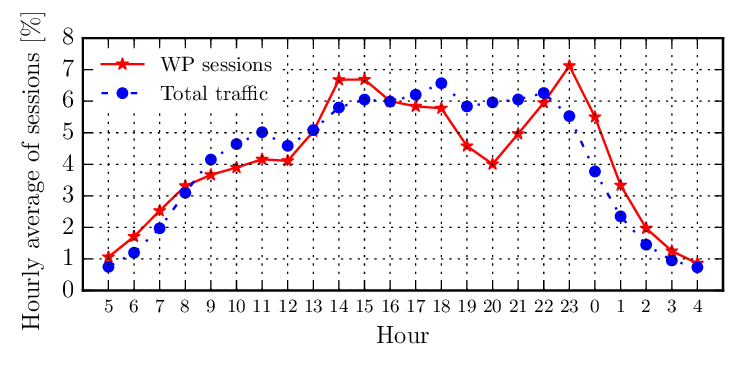}
    \caption{Average hourly percentage of number of \WP sessions and total traffic. }
    \label{fig:hour_of_day_porn_global_activity}
    \end{center}
\end{figure}

\begin{figure}[!t]
    \begin{center}
    \includegraphics[width=\textwidth]{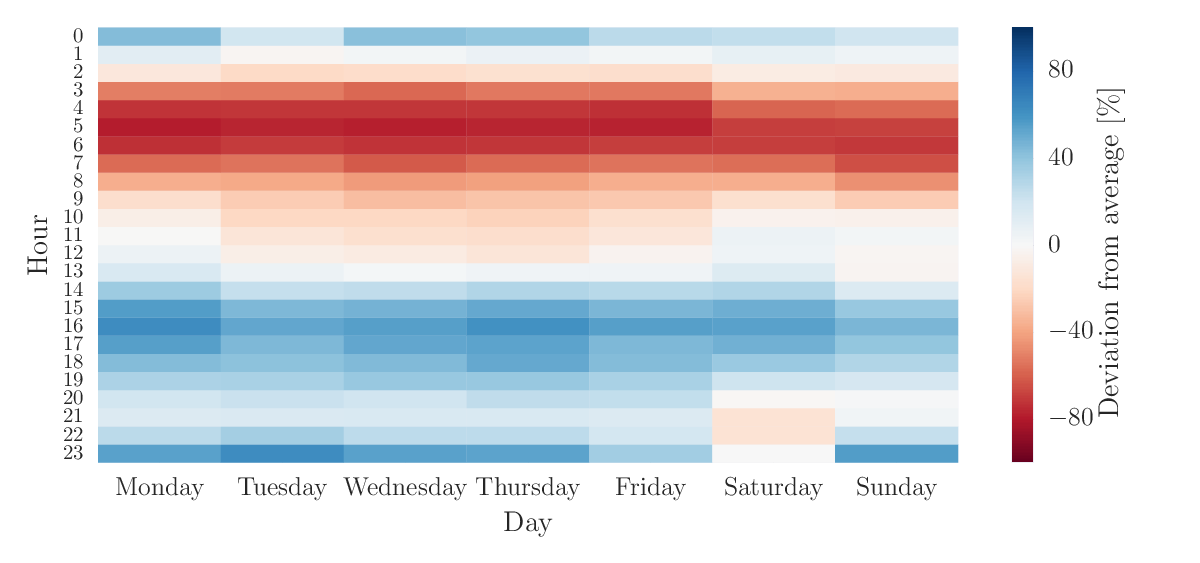}
    \caption{Weekly breakdown of hourly \WP usage. Heat-map of deviation from  hourly average. }
    \label{fig:hour_and_day_count}
    \end{center}
\end{figure}

\WP consumption also changes during different time of day. Fig.~\ref{fig:hour_of_day_porn_global_activity} provides the average percentage of sessions across the 24 hours of the day (red solid line). For ease of visualization, we start the $x$-axis from 4am, correspondent to the lowest value of the day. The two higher peaks are immediately after lunch time (2pm - 4pm) and after dinner (9pm - midnight). 
In addition to \WP traffic, the figure also reports the overall trend considering all HTTP transactions, regardless their nature (dashed blue line).
Comparing \WP to total traffic, some differences are noticeable; the peaks do not overlap, and the latter is definitely more balanced over daylight hours. An hypothesis for those divergences may be related to the fact that accessing pornographic websites is likely to be a private and leisure activity confined to intimate moments.
At a global scale, Pornhub service has found similar results.\footnote{See footnote \ref{note1}} 
The average session time reported by Pornhub for Italy is 9 minutes and 30 seconds, similar to what observed from our analysis. 
We also provide a breakdown across both hours and days of the week, with Fig.~\ref{fig:hour_and_day_count} showing the heat-map of the percentage variations from the gross weekly average (white color). Warmer tones register values below average, while colder ones show values above.  Notice some clear diminishing traffic on Saturday evening (7pm - midnight) and some increasing traffic on Saturday, Sunday and Monday morning (9am - 1pm). Indeed, many commercial activities are closed on Monday morning in the monitored country, perhaps influencing this behavior. Again, Pornhub data shows comparable results, with their heatmap having peaks of traffic in more or less the same time frames (2pm - 5pm) and (10pm - midnight).
Considering the cumulative daily accesses, Mondays register the highest values and Saturdays the lowest.


\begin{figure}[!t]
    \begin{center}
        \includegraphics[width=.75\columnwidth]{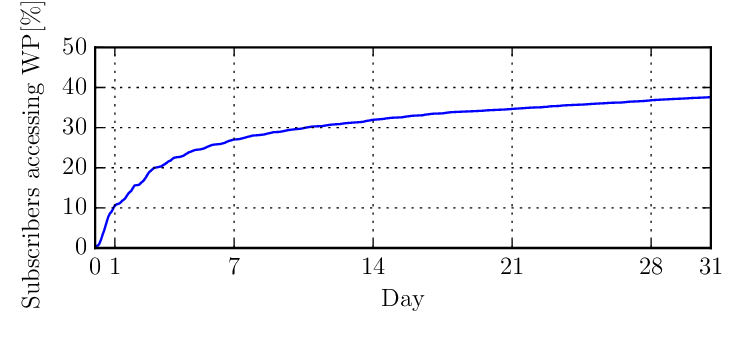}
        \caption{Cumulative percentage of subscriptions accessing \WP at different time in the trace.}
        \label{fig:user_discovery}
    \end{center}
\end{figure}

Finally, we provide an overall picture about the fraction of all monitored subscribers accessing \WP website. 
Although our dataset does not contain fine-grained details about \WP pervasiveness, we can still show the fraction of subscriptions where at least one user accessed \WP during our period of observation.
In Fig.~\ref{fig:user_discovery}, the $x$-axis represents the 31 days of our reference month (being day 1 October $1^{st}$, 2016 and day 31 October $31^{st}$, 2016), while $y$-axis reports the cumulative fraction of subscriptions that accessed at least one \WP website. Considering a single day, less than 12\% of subscriptions accessed \WP, but this fraction raises to 27\% after a week. At the end of the month it reaches 38\%, meaning that more than one subscription over three generated traffic toward \WP websites at least once in a month.
For comparison, YouTube and Netflix are daily accessed by 45\% and 3\% of subscribers respectively. Considering social networks, 60\% and 25\% of subscribers contact Facebook and Instagram on a daily basis, respectively.\footnote{A deeper analysis can be found in our previous work~\cite{tstat5years}.}

\subsection{\WP websites at a glance}
 
 \begin{figure}[!t]
    \begin{center}
        \includegraphics[width=.9\columnwidth]{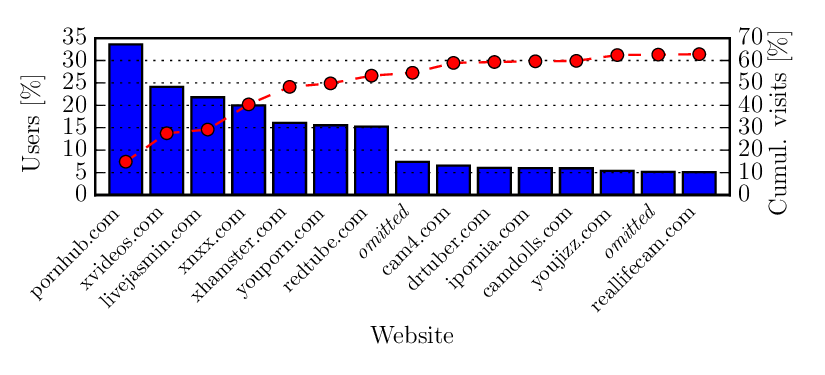}
        \caption{Top-15 \WP websites ranked according to percentage of users accessing them. Cumulative percentages of their visits with respect to all \WP visits are also shown. }
        \label{fig:domains}
    \end{center}
\end{figure}

In this section, we briefly describe \WP website popularity and pervasiveness.
Similarly to the Internet global trend, the market is dominated by few big players. Looking at the Alexa rank,\footnote{See footnote  \ref{notaAlexa}} three \WP websites appear among the top-50, namely \url{pornhub.com}, \url{xvideos.com} and \url{livejasmin.com}, with the first one ranked 29th, just behind \url{linkedin.com}. 
Considering our dataset, we observe a similar situation, with over-the-top companies leading the rank. In Fig.~\ref{fig:domains} we show the percentage of users reached by the top-15 \WP websites using bars (left-most $y$-axis), and the cumulative percentage of visits to these services (red line, right-most $y$-axis). In total users accessed $7\,048$ different websites during the entire month. The top-3 websites in our dataset match exactly Alexa rank, with \url{pornhub.com} being accessed by 34\% of the users. Global tendencies are reflected in our top-15, with only 2 \textit{omitted} websites as local representative of the monitored country. Considering the percentage of visited webpages, \url{pornhub.com} alone accounts for 14\% of them, and the top-15 together approximately 63\% of all \WP visits. The percentage reaches 90\% considering the top-204 websites, confirming the concentration of users around top services. Interestingly, very similar numbers hold for the overall traffic (including also non-\WP websites), with top-15 accounting for 61\% of traffic and 90\% due to 195 websites.


Finally, we notice that 3 out of 15 \WP websites of Fig.~\ref{fig:domains} belong to MindGeek, a company owning \url{pornhub.com}, \url{redtube.com}, \url{youporn.com}, and dozens of other websites.\footnote{\url{https://goo.gl/UgLqAj}} MindGeek websites account for more than 20\% of accesses in our dataset, making it a market leader. For comparison, the second website in terms of users and visits is \url{xvideos.com} (owned by WGCZ Holding), with less than half the users of MindGeek services, according to our data, suggesting a scenario where the ecosystem is lead by few big players in a dominant position.



\section{Conclusion}
\label{sec:conclusion}

In this paper we offered a quantitative analysis concerning web pornography consumption. To the best of our knowledge, we are the first to use network passive measurements to study the interactions of users with these services. We followed an exploratory approach on data, focusing on questions, topics and metrics typically 
analyzed in previous surveys and research works, \eg frequency of fruition and the time spent on \WP. We found interesting results, some typical of the observed population and others capable of confirming global trends.

Our results draw the attention to
a large and active group of users, and may be helpful for researchers that study web services consumption and human behavior at large. The obtained outcomes can be checked and verified, thanks to the fact that we release our anonymized dataset.
Furthermore, the chosen metrics allowed a comparison with outcomes of previously conducted surveys, and mostly confirmed their results.

\newpage

\bibliographystyle{splncs04}
\bibliography{paper}

\end{document}